\documentclass[sigconf]{acmart}
\AtBeginDocument{%
  \providecommand\BibTeX{{%
    \normalfont B\kern-0.5em{\scshape i\kern-0.25em b}\kern-0.8em\TeX}}}

\copyrightyear{2024}
\acmYear{2024}
\setcopyright{rightsretained}
\acmConference[CSCW Companion '24]{Companion of the 2024 Computer-Supported Cooperative Work and Social Computing}{November 9--13, 2024}{San Jose, Costa Rica}
\acmBooktitle{Companion of the 2024 Computer-Supported Cooperative Work and Social Computing (CSCW Companion '24), November 9--13, 2024, San Jose, Costa Rica}
\acmDOI{10.1145/3678884.3681895}   
\acmISBN{979-8-4007-1114-5/24/11}

\usepackage{booktabs} % for better looking tables

\usepackage{float} % for the storyboard table
\usepackage{placeins} % appendix formatting?

\begin{document}


% The title of your work should use capital letters appropriately -
% \url{https://capitalizemytitle.com/} has useful rules for
% capitalization. Use the {\verb|title|} command to define the title of
% your work. If your work has a subtitle, define it with the
% {\verb|subtitle|} command.  Do not insert line breaks in your title.

% If your title is lengthy, you must define a short version to be used
% in the page headers, to prevent overlapping text. The \verb|title|
% command has a ``short title'' parameter:
% \begin{verbatim}
%   \title[short title]{full title}
% \end{verbatim}

\title[Generative AI in Small Business]{Exploring the Role of Social Support when Integrating Generative AI in Small Business Workflows}
%Can Local Networks Address Disparities in Use of Generative AI Technologies among Small Business Owners?

%
% The "author" command and its associated commands are used to define
% the authors and their affiliations.
% Of note is the shared affiliation of the first two authors, and the
% "authornote" and "authornotemark" commands
% used to denote shared contribution to the research.
\author{Quentin Romero Lauro}
\orcid{0009-0000-0620-2146}
\affiliation{
    \institution{University of Pittsburgh}
    \city{Pittsburgh}
    \state{Pennsylvania}
    \country{USA}
}
\email{quentinrl@pitt.edu}

\author{Jeffrey P. Bigham}
\orcid{0000-0002-2072-0625}
\affiliation{
    \institution{Carnegie Mellon University}
    \city{Pittsburgh}
    \state{Pennsylvania}
    \country{USA}
}
\email{jbigham@cs.cmu.edu}

\author{Yasmine Kotturi}
\orcid{0000-0001-6201-7922}
\affiliation{
    % \institution{\mbox{University of Maryland, Baltimore County}}
    \institution{\mbox{University of Maryland, Baltimore County}}
    \city{Baltimore}
    \state{Maryland}
    \country{USA}
}
\email{kotturi@umbc.edu}

\renewcommand{\shortauthors}{}

\begin{abstract}
  % Abstract

%, magnified by the rapid pace of technological advancements.
% Using generative AI for small business comes with a set of risks such as legal and ethical considerations, magnified by the rapid pace of technological advancements.

Small business owners stand to benefit from generative AI technologies due to limited resources, yet they must navigate increasing legal and ethical risks.
In this paper, we interview 11 entrepreneurs and support personnel to investigate existing practices of how entrepreneurs integrate generative AI technologies into their business workflows.
Specifically, we build on scholarship in HCI which emphasizes the role of small, offline networks in supporting entrepreneurs’ technology maintenance.
We detail how entrepreneurs resourcefully leveraged their local networks to discover new use-cases of generative AI (e.g., by sharing accounts), assuage heightened techno-anxieties (e.g., by recruiting trusted confidants), overcome barriers to sustained use (e.g., by receiving wrap-around support), and establish boundaries of use. 
Further, we suggest how generative AI platforms may be redesigned to better support entrepreneurs, such as by taking into account the benefits and tensions of use in a social context.

\end{abstract}







\begin{teaserfigure}
  \includegraphics[width=\textwidth]{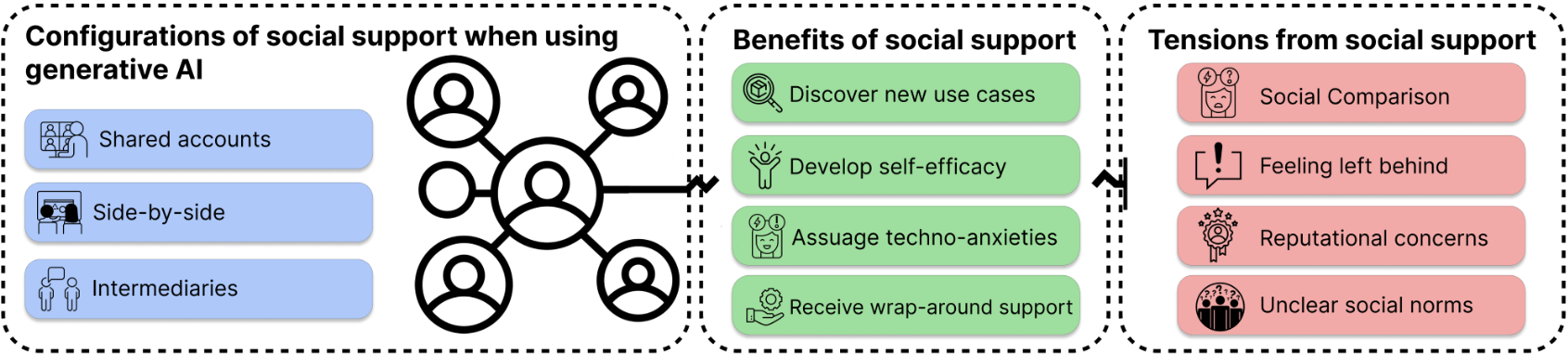}
  \caption{
    % Our interview study probed small business owners' existing practices and desired practices for how to integrate generative AI into their workflows. Specifically, given entrepreneurs' tendency to seek social support when using technology for their business, we detail the importance of \textbf{shared accounts} (e.g. collaboratively prompting by sharing ChatGPT premium logins, \textbf{side-by-side} (e.g. co-articulating prompts with a technically savvy partner or support personnel, and through \textbf{intermediaries} (e.g. a partner translates the entrepreneur's struggle wording an email into a use case of ChatGPT and facilitates use). 
  Through an interview study, we detailed how entrepreneurs used their social support networks to integrate and maintain generative AI technologies in their business workflows (e.g., shared accounts with co-owners, side-by-side tutoring sessions, proxy use through trusted intermediaries); we then probed the benefits and tensions of social support. %when navigating a novel technology 
  % allow entrepreneurs to discover new use cases of generative AI, develop self-efficacy in using  the technology, assuage techno-anxieties, and receive wrap around support when using the technology. Despite the benefits of social support, we outline tensions from these configurations: demoralizing social comparison around use, feelings of being left behind, reputational concerns, and unclear social norms arising from disparate expectations of others when using generative AI together.
  % }
  }
  \label{fig:teaser}
  \Description{Forms, Benefits, and Tensions of Social Support in Using Generative AI. Different configurations of social support when using generative AI include shared accounts, side-by-side use, and use via intermediaries. While, benefits of social support include discovering new use cases, developing self-efficacy, easing techno-anxieties, and receiving wrap-around support, social support when using generative AI comes with tensions including social comparison, feeling left behind, reputational concerns, and unclear social norms.}
\end{teaserfigure}

% Different forms of social support when using generative AI include...
% Different benefits entreprneuers received from these forms of social support included 
% If there is a higher level point, that one may get by having this in front of them, state that at the end.

% In this paper, we describe local entreprenuers 


\maketitle

% --------------------------------
\section{Introduction}
Small business owners who effectively leverage generative AI for their business can boost their efficiency, creativity, and ultimately increase their bottom line \cite{Otis2023UnevenImpact}.
But integrating these novel technolog-ies---machine learning models that can generate high-quality text, images, and other content from training data \cite{IBMGenerativeAI}---into existing workflows might not be as straightforward as generative AI platforms tout \cite{openai_chatgpt}.
For instance, entrepreneurs must consider the implications of providing intellectual property to generative AI platforms, and the ill-defined legality and ethics of using output generated from mining the web \cite{nytimes2023Sue}.
In addition, prompt engineering, an essential skill to use generative AI platforms such as ChatGPT and DALL-E, is not as intuitive as it seems and non-expert users struggle to iterate on prompts to generate desired output~\cite{10.1145/3544548.3581388}. 
These challenges are further heightened for entrepreneurs from resource-constrained communities, or lean economies \cite{dillahunt2018entrepreneurship}, who must deploy resourceful tactics to overcome limited access to infrastructural support such as education and  equipment~\cite{avle2019additional}.
% In fact, small business owners who benefit from generative AI tend to be more successful; in other words, the rich get richer~\cite{Otis2023UnevenImpact}.

In response to entrepreneurs' difficulties with keeping pace with technology advancements, human-computer interaction (HCI) scholars have detailed at length the role that small, offline networks play in helping small business owners plan and implement a strategy for tech maintenance~\cite{10.1145/3313831.3376363, dillahunt2018entrepreneurship, 10.1145/3491102.3517708}. 
These networks, or ``collectives,'' comprise peers, mentors, friends, and even loyal customers, and play an especially important role for entrepreneurs from lean economies because of sparse access to infrastructural support for technology adoption~\cite{10.1145/3313831.3376363}. 
For instance, local community groups can provide vetted information on which platforms and technologies entrepreneurs should adopt~\cite{kotturi2024deconstructing, dillahunt2018entrepreneurship}, skill training opportunities~\cite{10.1145/3491102.3517708}, as well as general encouragement (critical, as many entrepreneurs work in isolation)~\cite{dillahunt2018entrepreneurship, 10.1145/3313831.3376363}. 
Yet, it remains unclear how these offline networks may facilitate entrepreneurs' adoption of generative AI technologies into their business. 
While recent work suggests that local community centers may play an essential role in entrepreneurs' introduction to generative AI through guided, communal introduction~\cite{kotturi2024deconstructing}, not all entrepreneurs have reliable access to such spaces.
In addition, the rapid pace of change---with new generative AI platforms and features deployed every day---makes it hard for even the most advanced business incubators to update curricula.
Finally, entrepreneurs may have different preferences of use of generative AI given the multitude of unfolding ethical, legal, and inaccuracy concerns as use relates to their business (and therefore their livelihood)~\cite{kotturi2024deconstructing, mitsloan-generative-ai}.

This paper, therefore, investigates how small business owners use various forms of social support from their offline, informal networks to determine how to integrate generative AI into their business workflows.
% In particular, we focus on the offline social support which entrepreneurs from lean economies often rely on for originating from the small, offline networks which are essential for
We ask: \textbf{how does social support facilitate entrepreneurs' use (and non-use) of generative AI technologies?}
We interviewed 11 entrepreneurs and entrepreneurship support personnel (e.g., tech support staff at local community centers) to better understand existing practices for how entrepreneurs integrated generative AI, as well as how they used their offline networks to guide these decisions, support maintenance and repair, and refusal of use (e.g. deciding to not use generative AI~\cite{garcia2020no}). 
While not ``high-tech entrepreneurs''~\cite{roberts1991entrepreneurs}, the entrepreneurs in our study used myriad computing tools to market, create, and sell their goods~\cite{10.1145/3491102.3517708}, and primarily targeted their local economy with their product or service-based businesses (i.e., local entrepreneurs~\cite{hui2018making}).
For instance, the types entrepreneurship represented in our study included running a streetwear clothing brand, offering gifts and party planning, providing choreography and dance classes, running letterpress and educational programming, and selling paintings; see Table~\ref{tab:participants}.
To engage entrepreneurs who were not yet embedded in a network with expertise on generative AI, interviews included a series of probes (in the form of paper storyboards~\cite{storyboardingPractices}) to enable deeper discussions for how to design generative AI systems with such a social context in mind. 
Combining semi-structured interviews and probes, therefore, enabled a deeper discussion of both current and desired states for how social support may---and may not---be helpful when integrating generative AI technologies into business workflows.

We found that offline support networks' role ranged from providing supportive nudges to overcome the associated anxieties with getting started, to a collaborative partner in formulating prompts and finding out new ideas for how to apply these nascent technologies to small business.
For example, despite tools' single-user assumption (e.g., standard accounts on ChatGPT presume only one user), entrepreneurs appropriated existing systems for shared use (i.e., by creating one account and sharing passwords with multiple users) to overcome operational barriers, like prompt engineering.
% to using generative AI and to build self-efficacy through encouraging others and receiving encouragement.
These networks---comprising business incubators and accelerators, community centers, close friends, spouses, and business partners---served as sources of discovery over time, showcasing the diverse applications of generative AI, and as mediators, helping to refine requests and ensure useful outputs. 
We observed that this shared use was critical to keep up with the pace of innovation: entrepreneurs who sought continuous support through their local networks reflected on how they observed how others used the technology, specifically those whom they trusted with their business affairs. 
% affords tended to have more sustained use of generative AI beyond introductions.
Additionally, participants in our study reflected on their boundaries when adopting generative AI, which were sometimes informed by their networks' attitudes towards the nascent technology (e.g., refusing to use generative AI when it might incur a reputational cost attributed to duplicity).
%(e.g., refusing to use generative AI to create an artist statement for the fear of others' perceptions)
% Sometimes boundaries were informed by budding social norms around using generated content for certain tasks, like an artist statement.}
% specifying ethical, and sometimes socially informed, boundaries of use.
% For example, a peer's non-use of generative AI sometimes informed an entrepreneurs own non-use.
% Additionally, we observed that local entrepreneurs established boundaries 
Further, we observed tensions that entrepreneurs experienced when collaboratively using generative AI platforms, and detailed how entrepreneurs' considered the emergent social norms when sharing generative AI technologies.

Taken together, this paper contributes empirical data on how small business owners integrated generative AI technologies into their workflows and how their supportive networks facilitated this process. 
We consider how generative AI technologies may be redesigned to support the social context of use among small business owners, such as by taking into account the unique nature of entrepreneur relationships.
For instance, we posit how systems may take into account the mix of cooperation and competition among entrepreneurs (i.e., ``coopetition''~\cite{kuhn2015little}), when fostering critical use in a social context.

% systems can empowers users to decide \textbf{who} to engage in generative AI use with and dynamic controls for information sharing to provide entrepreneurs more agency on \textbf{what} is shared when engaging with generative AI in a social context.
% For example, by giving entrepreneurs more agency on \textbf{who} and \textbf{what} to share when engaging generative AI collaboratively.

% \textcolor{blue}{TODO one sentence on concrete design implication based on coopetitive nature of relations.}
%By accounting for the coopetitive nature and knowing when socail use may be more or less desirable, generative AI technologies can...
\section{Related Work}

\subsection{Entrepreneurship in the Digital Age}
Entrepreneurship in the digital age can be a driver for innovation, economic mobility, and social change~\cite{10.7551/mitpress/2333.001.0001}, especially due to reduced startup costs and widening customer reach resulting from online, e-commerce, and mobile platforms~\cite{digital-entrepreneurship}.
Yet, scholars have questioned whether digital entrepreneurship is truly a ``democratizing'' force~\cite{10.7551/mitpress/2333.001.0001}, surfacing how the constant self-upgrading and technical maintenance required for entrepreneurs to stay relevant disproportionately disadvantages under-resourced and non-technical entrepreneurs~\cite{avle2019additional, hui2018making}.
% due to growing digital divides \cite{bbc-digital-literacy}, 

In particular, prior scholarship in HCI has outlined an array of technology challenges that local entrepreneurs---entrepreneurs who target their local economy to overcome limited job opportunities~\cite{hui2018making}---face, such as heightened risk given resource constraints, lower social capital, and infrastructural barriers to access~\cite{dillahunt2018entrepreneurship, avle2019additional, 10.1145/3313831.3376363, hui2018making, 10.1145/3313831.3376587, kotturi2024peerdea}. 
In light of these challenges, scholars have highlighted the crucial role of small, local networks for entrepreneurs in resource-constrained communities to overcome barriers in technology adoption and economic mobility~\cite{10.1145/3491102.3501949, 10.1145/3313831.3376363, 10.1145/3491102.3517708}.
For instance, Hui \textit{et al.} found how local entrepreneurs prefer to stay digitally engaged by leveraging low-tech social support such as in-person meetings, paper planning tools, and resource-connecting organizations \cite{10.1145/3313831.3376363}.
Dillahunt \textit{et al.} presented ``the Village'' model of mentorship, and found that in-person interaction and trust, which technologically-mediated relationships may not facilitate, were required to sustain economic mobility.
This importance of social support extends to the context of generative AI.
For instance, introducing generative AI in a communal setting among local entrepreneurs helped to assuage the anxieties which these nascent and powerful technologies can elicit upon introduction~\cite{kotturi2024deconstructing}, sometimes referred to as ``techno-anxieties''~\cite{technoAnxiety}.
However, not all entrepreneurs have reliable access to community spaces, and must instead rely on more informal support networks such as close friends and family for social support.
We, therefore, build on this scholarship to investigate how small business owners leverage their informal, social support networks to determine how to integrate generative AI into business workflows, and the benefits and tensions with such an approach.

% \vspace{-.4cm}
\subsection{Generative AI in HCI}
Recent advancement in large-language and diffusion models have stimulated many recent end-user developments of generative AI technologies.
Interaction with generative AI platforms has mainly been centered around prompt engineering, in which a user enters text to prime the system to generate an output (text, image, video, speech) \cite{McKinseyPromptEngineering}.
Despite its advertised simplicity, recent work with non-experts, even those from technical backgrounds, highlight the challenges when using prompting as an interaction technique with large-language models (LLMs)~\cite{10.1145/3544548.3581388, kotturi2024deconstructing}. 
Given the limitations of prompting, HCI researchers and practitioners have tried to address these shortcomings in a number of ways. 
For instance, researchers have created and published guidelines and practices for prompting generative systems \cite{10.1145/3491102.3501825}, investigated how to aid in the prompt writing process by automatically generating variations of prompts \cite{9908590}, and enabled users to repeatedly assess and iterate on prompts~\cite{10.1145/3581641.3584059}.
% Moving away from prompting, other work has investigated multi-modal techniques for more expressive interaction and exploration of generative AI systems capabilities \cite{10.1145/3544549.3577043}. 
%Some applications seek to narrow the scope of systems' generalizability and investigate how generative AI can be utilized with more targeted interaction techniques (e.g. sliders \cite{10.1145/3313831.3376739}) to support specific tasks, such as music creation \cite{10.1145/3313831.3376739} and creative and scientific writing \cite{10.1145/3544548.3580782, 10.1145/3491101.3519873, 10.1145/3532106.3533533}. 
% While past scholarship moves towards different interaction techniques that may lower operational barriers to using generative AI, like prompting, our work investigates these how local entrepreneurs utilize their local networks in overcoming these challenges while using generative AI to support their business.
Further, recent work has shown how socially-situated generative AI can aid co-creation and collaboration between designers \cite{10.1145/3544549.3585680} and evidenced the benefit of generative AI in stimulating collaborative ideation processes \cite{10.1145/3544549.3573802}. 
While prior scholarship focused on settings where users are working towards shared goals, our work considers the use of generative AI in a context where users seeking social support exhibit behavior as both collaborators and competitors \cite{kuhn2015little}, and additionally must navigate a complex environment for information sharing to safeguard intellectual property and reputation.

\section{Methodology}
% We conducted 11 semi-structured interviews with seven entrepreneurs and four entrepreneur support staff.
We conducted 11 semi-structured interview comprising both open-ended interview (Section~\ref{sec:interviews}) and probe-based questions (Section~\ref{sec:storyboards}) with seven entrepreneurs (E1-E7) and four entrepreneur support staff (P1-P4); see Table~\ref{tab:participants}.
We recruited participants from two local entrepreneurial hubs based in Pittsburgh focused on racial and gender equity in entrepreneurship and technology.
Three of four support personnel held or were pursuing graduate degrees, and provided technical support within the entrepreneurial hubs (e.g., as one-on-one tutors~\cite{10.1145/3491102.3517708}).
The research team had multi-year collaborations with both entrepreneurial hubs, where the last author started an on-going technology support program and entrepreneurial programming in both hubs~\cite{10.1145/3491102.3517708, kotturi2024peerdea}, providing the needed foundation of mutual understanding and trust, and historical context to engage in the research study~\cite{le2015strangers, harrington2019deconstructing}.  
Entrepreneurs had a range of product and service-based businesses such as apparel, dance, gift baskets, painting, and education. 
Recruitment prioritized those who had some exposure or experience with generative AI platforms such as: Canva text-to-image, ChatGPT, DALL-E, or Gemini.
Details on entrepreneurs' businesses and uses of generative AI can be found in Table~\ref{tab:participants}.
Participants were compensated \$20 per hour.

% The support staff were involved with entrepreneurship mentoring in various capacities, and entrepreneurs had a range of businesses 

\subsection{Interviews}
Interviews included two parts: open-ended questions about participants' backgrounds and businesses, followed by probing potential futures with paper storyboards.

\subsubsection{Part 1: Background questions}
\label{sec:interviews}
First, we asked a series of questions about participants' businesses and/or experiences mentoring business owners, their technology background or how they have integrated technology within their business (e.g., ``How comfortable are you with technology in your daily operations?''), and what their support networks look like (e.g., ``What kind of support do you find helpful when navigating problems in your business?'').
We also asked questions specifically about generative AI technologies: which platforms participants used and their concerns. 
As interviews were rooted in past experiences, we also wanted engage participants in deeper discussion about potential and preferred futures, as it related to integrating generative AI technologies into their business workflows, and how their support networks (existing or desired) might play a role in this next technological frontier.
Therefore, in the second part of interviews, we used probes in the form of paper storyboards, detailed in the next section.

% =========== STORYBOARD FIGURE ===========
% \begin{figure}[t]
% \centering
% \begin{minipage}{0.45\textwidth}
%   \centering
%   \includegraphics[width=\textwidth]{Appendix Content/storyboard_1.png}
%   % \caption{Storyboard 25}
%   \label{fig:appendixImage1}
% \end{minipage}
% \hspace{0.05\textwidth}
% \vspace{-.78cm}
% \caption{We used paper storyboards to probe entrepreneurs on potential and preferred futures for how social support may be more and less helpful when integrating generative AI to their small business (See Supplemental Materials for storyboard subset).}
% \vspace{-.64cm}
% \label{fig:storyboards}
% \end{figure}
% ==========================================

\subsubsection{Part 2: Storyboards as Probes}
\label{sec:storyboards}
Storyboards---or short scenarios of possible futures---can be an effective way to engage users in grounded discussions of preferred futures~\cite{storyboardingPractices}, especially in a workplace context undergoing rapid technological transformation~\cite{kawakami2023sensing}.
In our study, we created a set of paper storyboards to probe how entrepreneurs may (and may not) integrate generative AI into their business workflows, and how their offline social networks may facilitate this process.
To derive storyboard scenarios, we used \textit{Ideation Decks}, a brainstorming tool that helps define specific design problems within a broader problem space~\cite{golembewski2010ideation}. 
To ensure storyboards represented adequate coverage of the relevant design space and was informed by prior work, we generated five \textit{category suits}, with 5-11 \textit{instances} in each category: (1) which local network the entrepreneur sought support from such as local community centers~\cite{10.1145/3491102.3501949}, peers~\cite{kotturi2024peerdea}, technical providers~\cite{10.1145/3491102.3517708}, family~\cite{10.1145/3313831.3376363}  (2) the form factor for delivering social support when not co-located such as chat-based, email-based, browser-based, or SMS interventions (to meet entrepreneurs where they are, what tools they currently use~\cite{kotturi2024deconstructing}, rather than introducing an entirely new system~\cite{piggybackPrototyping}) (3) the stages of the entrepreneurs' workflows where generative AI may be more or less helpful, such as early on when brainstorming new product or service ideas~\cite{gero2022sparks} versus when refining materials such as business plans (4) various pain points for entrepreneurs as detailed in prior work~\cite{10.1145/3491102.3517708}, such as generating content (for social media, websites, etc)~\cite{10.1145/3491102.3517708}, or navigating legal issues when using AI-generated content for business purposes~\cite{kotturi2024deconstructing}, and finally, (5) various configurations of blended human-machine support given that entrepreneur support networks may not always be available (especially busy peers~\cite{kotturi2024peerdea}). 
Note: each storyboard did not contain all category suits, but rather a compact subset in order for the scenario to fit within a short, 6-panel layout.
See the Appendix~\ref{appendix:A} for a subset of storyboards created.

% Through an iterative and reflexive process within the research team, the first author drafted 20 storyboards prior to commencing interviews.
Throughout the duration of the study and based on participants' feedback, we removed, refined, and added storyboards to maximize discussion, resulting in 26 storyboards total.
% Ultimately, of the 19 of the 26 storyboards (S1-S26) generated were presented to participants in our study. 
Language in the presented storyboards avoided technical jargon for non-experts.
Following background questions (Part 1), the interviewer presented a subset of the storyboards that best responded to the emergent needs that arose from participants' responses to background questions. 
For example, if an entrepreneur was not interested in using image-based generative AI tools, we did not show them storyboards which featured a scenario of an entrepreneur wanting to use image-generation technology. 
% Even so, we ensured our selection of storyboards provided sufficient variation in concepts, such that storyboards did not cluster around a single idea. 
After storyboards were presented, the interviewer asked participants to reflect upon and compare them.

% \vspace{-.4cm}

% All support personnel had exposure to generative AI, though some preferred to limit use because of ethical, efficacy, and efficiency concerns. 
% The presentation of storyboards was adapted to fit the local community context in which the participant was in.
% for the duration of the interview, and additional compensation to cover the cost of transportation if the participant traveled to [University] for the interview.
% participant table
\begin{table*}[t]
\centering
\small
% \rowcolors{2}{gray!15}{white} % Alternating row colors
\begin{tabular}{p{0.2cm} p{2.2cm} p{6.43cm} p{6.2cm} p{1cm}}
\toprule
% \rowcolor{gray!25}
\textbf{PID} & \textbf{Role} & \textbf{Job/Business Description} & \textbf{Uses of Generative AI} & \textbf{Frequency} \\
\midrule
E1 & Local Entrepreneur & Streetwear clothing brand & Creating animation code, podcast editing, captioning & Often \\ 
% E2 & Local Entrepreneur & Gift Basket Business & Images and descriptions of gift baskets for social media & Daily \\
E2 & Local Entrepreneur & Gift Basket Business & Image \& text generation for social media & Often \\ 
E3 & Local Entrepreneur & Choreography, dance classes and fitness lessons \& podcast & Background \& financial information & Minimal \\ 
% E4 & Local Entrepreneur & Wholesale letterpress  \& educational programming for youth & Organizing lesson plans & Less than 1/month \\ 
E4 & Local Entrepreneur & Wholesale letterpress  \& educational programming & Organizing lesson plans & Minimal \\ 
E5 & Local Entrepreneur & Bolly-fusion dance classes & Writing emails & Often \\ 
% E6 & Local Entrepreneur & Wholesale letter press and stationary company, also providing educational programming for youth & Rewording copy, Generating captions for Instagram posts & 1/week\\ 
E6 & Local Entrepreneur & Wholesale letterpress \& educational programming & Rewording copy, Generating social media captions & Minimal \\ 
E7 & Local Entrepreneur & Artist; Painting & Creating Instagram Captions & Minimal \\ 
% \addlinespace
P1 & Support Personnel & Tech-support volunteer & --- & N/A \\ 
P2 & Support Personnel & Tech-support volunteer & Cooking recipes & Minimal \\ 
% P3 & Community Provider & Co-founder, local entrepreneurship hub focused on gender and racial equity through tech access, community workshops, and peer-support & --- & --- \\
P3 & Support Personnel & Co-founder of local entrepreneurship hub & Coding & Minimal \\
P4 & Support Personnel & Co-founder of local entrepreneurship hub & Writing emails & Often \\ 
\bottomrule
\end{tabular}
\caption{Participants included both entrepreneurs (``E\_'') and support personnel (``P\_'') who had diverse businesses, used generative AI for a range of tasks and at various frequencies; together, this provided well-rounded perspectives in this exploratory study.}
\label{tab:participants}
\Description{Summary of Participants' Roles and AI Usage. This table summarizes participants, including local entrepreneurs and support personnel, detailing their roles, business descriptions, uses of generative AI, and the frequency of AI usage. Entrepreneurs used generative AI for tasks such as animation code creation, podcast editing, captioning, and social media content generation. Support personnel primarily provided tech support and engaged in activities like coding and email writing.}
\end{table*}

\subsection{Data Analysis}
On average, interviews lasted 47.6 minutes and 5-10 storyboards were presented in each interview. 
5 out of 11 interviews were conducted remotely via Zoom, based on participants' preferences.
In-person interviews took place at Carnegie Mellon's laboratory or a local community center. 
% Before each interview, the interviewer gave an overview of the study and emphasized consent. 
% During each interview participants were first asked a series of background questions regarding their business or experience as support provider, comfort with technology, and experience with generative AI.
% For example: ``Can you tell me about your business and what you do?'', ``Are there any tasks that you find tedious or challenging that you wish could be easier?'', ``What emotion comes to mind when you hear the term AI?''.
Both the first and last authors took turns interviewing participants while the other took field notes; all interviews were audio recorded.
Audio recordings were transcribed automatically with Temi.com, which the first author then checked for errors and fixed by hand before the team engaged in thematic analysis.
We adhered to quote editing conventions commonplace in applied social science research methods \cite{quoting-convention}.
The first author created an affinity diagram, clustering key themes which emerged from open-coding the interview transcripts such as tensions to use and social support (operational barriers, privacy concerns) and benefits of social support (staying in the loop, reducing costs). % as well as forms of social support (in-person, side-by-side, intermediaries).

The first author then wrote analytic memos for each participant with an average word count of 1,060 words~\cite[Page 72]{charmaz}.
Analytic memos---or short narratives used to translate rich qualitative data into narrative format---were critical to triangulate participants' responses to interview questions, alongside open-ended discussions about the storyboards they were presented~\cite{charmaz}.
Memos were structured around sections which expanded upon the benefits and tensions of social support: what AI technologies participants used, desired to use, or refused to use; social support preferences; workflows.
Each analytic memo was then shared with the research team for review, and the last author provided detailed comments to engage in back and forth discussion.

\section{Findings}
The entrepreneurs in our study leveraged their offline networks to appropriate existing single-user generative AI platforms for shared use by sharing accounts and working together to use generative AI collaboratively in-person and online.
Entrepreneurs reflected on the value of these small networks, by either reflecting on their own experiences or in response to brainstorming with the probes if they had not yet had the opportunity to build up such a network. 
For instance, in these small, trusted networks, entrepreneurs overcame technology-driven anxiety, collaborated on formulating prompts, and discovered new ways to use the technology.
However, entrepreneurs had to navigate unclear standards for behavior, the potential for dampening self-efficacy, and embarrassment that could occur when collaborating with their network on generative AI platforms. 
To begin to unpack the dynamics the social networks facilitating entrepreneurs' use of generative AI in our study, we detail them here.

% =======================================================================
\subsection{Forms of Social Support, and Benefits}
Often, entrepreneurs' introduction to generative AI technologies were driven by a fear of being \textit{``left behind''} (P4). 
Entrepreneurs' offline networks were critical in turning this anxiety into action, by providing the supportive nudge to learn more about the technology. 
For example, E3, who ran a dance business and was also a full-time writer, described a conversation she had with a her spouse where she voiced her concern: \textit{``[If] Gmail autofills and completes my emails at this point, [what's] stopping it from auto-filling all of my documents that I'm spending ...three months writing?''}
% E3 also took this concern to her dance business. She wondered how AI might impact the future of choreography if trained off dance videos online.
She recalled her spouse's response, \textit{`` `You should just learn it and then be more informed and be less afraid.' ''}. 
E3 shared that this conversation with her spouse motivated her to join a Udemy class on prompting and start exploring ChatGPT more. 
E4 and E6, who owned a letterpress business together, encountered ChatGPT for the first time at a recent small business incubator. 
E4 described how her and her business partner (E6)---who were in their 30s---felt like \textit{``boomers,''} having not encountered the technology while everyone else seemed to be using it. 
Amidst this anxiety, E6 noted the helpfulness of another entrepreneur there who went on to teach her how to use ChatGPT.

Support personnel also took on an active role to critically engage entrepreneurs with the technology.
For instance, P4, the co-founder of a local entrepreneurial hub focused on equitable community development, felt the responsibility to be the person in entrepreneurs' networks that provided the supportive nudge for entrepreneurs to get started. 
P4 referred to the developments in the last year the \textit{``GPT Revolution''} where \textit{``It felt like [an] earthquake and a divide in the ground, and I was slowly moving away from the rest of society.''} 
They felt that if they did not \textit{``start integrating AI into [their] daily practices''} they would be \textit{``left behind''}. Because of this feeling, they took on responsibility to encourage the entrepreneurs that they serve to follow suit: \textit{``I tell everyone I work with that [they] really need to get on board because they will be left behind...Even if I don't necessarily feel that I'm on the other side of the divide yet, I certainly encourage the entrepreneurs and the other contractors and freelancers [and] business folks that I work with.''}
To P4, the ``divide'' represented the digital divide between those who are learning about generative AI and those who are not. 
In fact, E2 later learned about generative AI technologies from a workshop at the community hub that P4 co-founded.
Whether through peers, spouses, business partners or support personnel, entrepreneurs' local networks were key in engaging them in critical use.
% ==============================================================================

% ==============================================================================
% \subsection{The Social Context of Generative AI Use for Local Entrepreneurs
Despite predominant generative AI platforms being designed for single-user use, E1, E2, and E5 leveraged their local networks to collaboratively use generative AI technologies by sharing accounts and working together in-person. 
By working together while using generative AI technologies, entrepreneurs were able to observe prompting strategies, build self-efficacy, and discover new ways to apply the technology.
For instance, E1, who ran a streetwear clothing brand with six other business partners, shared access to a ChatGPT Premium account with everyone in the business. 
E1 noted that by sharing access to the account he gets the benefit of learning from his peers mistakes, \textit{``I only have to [fail] maybe once or twice because I've seen six other people try and fail in different ways.''}. %And I can see like, oh, does that work? No, it doesn't work. ...So without even doing it, I can see if [prompt] worked ...and then I can go off of that.
\textit{``Failing''} was when the prompt he entered did not produce the output that he had in mind. 
By seeing the prompts and the results of his peers, E1 could passively learn about what did and did not work well when prompting ChatGPT without the added time to figure it out for himself. 
% Further, even if a prompt was not a \textit{``failure''}, E1 shares how he could expand off of what others had already done. 
He called this \textit{``Group Work for AI''}.
% , where he can fail, learn, and iterate upon his business partners interaction with ChatGPT.
E1 also pointed out that by sharing an account, it opened a new way for him and his business partners to keep tabs on how certain projects were progressing. 

% E2, who operates a gift basket and party planning business, experienced shared, in-person use during a recent community workshop which partnered entrepreneurs with a technical expert for side-by-side support.
E2, who ran a gift basket and party planning business, described how she recently took part in a workshop on how to use generative AI. 
At the event she described how she was paired with a technical provider for side-by-side support while using ChatGPT.
This format helped E2 overcome the apprehension she had about using generative AI, \textit{``When we were able to sit around the table and you got to sit with a [provider] ... [who] makes you feel like you can do it yourself, and it makes you feel at ease.''}
For E2, who described her first experience with text and image generating technologies as \textit{``horrifying''}, using the technology alongside a technical provider helped her build self-efficacy and overcome her initial anxiety in getting started. 
% Another benefit E2 pointed out of this side-by-side, shared use of generative AI was that she could receive continuous support from start to finish: from logging into DALL-E, to co-articulating prompts.
When the workshop ended, E2 replicated side-by-side support while using generative AI with her spouse, who sometimes created prompts for her business tasks. 
When her spouse or tech providers were not available to meet side-by-side, E2 shared that she texts them \textit{``three or four times a day''} to get support.
E5, who ran a dance business and shared an account with her spouse, described using ChatGPT together: he brainstormed a new use case for ChatGPT and then created an initial prompt, acting as a translator of her needs and an intermediary to using the technology while she focused on daily business operations.
% In this way, E5 does not experience the friction of figuring out a use case for generative AI or prompting the system to elicit a helpful response.
While E5 described using ChatGPT very frequently, she has only used it on her own twice; the rest of the time, her spouse was there to translate her needs and mediate her use.
%she is still able to use the technology to refine emails for her business. 
%\textit{``twice the whole time ChatGPT has been a thing,''} 

While not all entrepreneurs in our study collaborated with others in their network to use generative AI platforms (because they did not have the network to do so, or preferred to work independent of others), when responding to storyboards that featured different forms of social support with generative AI (e.g. multi-user interfaces), entrepreneurs and support personnel pointed out the potential benefits.
For instance, in response to S17 (see Appendix Figure ~\ref{fig:storyboard_17}), which featured a SMS prompt sharing service for familiar entrepreneurs, E3 shared how she could work towards \textit{``developing [her] own skill set by emulating or using other folks as a starting point'' }---similar to how E1 learned from his business partners prompts through his shared ChatGPT Premium account.
Reacting to S9 (see Appendix Figure \ref{fig:storyboard_9}), which featured entrepreneurs with a shared community context pairing up to use a generative AI system, P2 and P3 reflected on how such pairings could \textit{``bridge the gap [of] comfort around these tools''} (P2)---similar to how E2 experienced in her use of ChatGPT at a recent community workshop.
Overall, entrepreneurs shared interest in systems that would further enable them to use generative AI technologies with a trusted confidant, with the hope of experiencing the support E1, E2, and E5 described to build self-efficacy and to discover new use-cases.

% in heightening questions of ``self-efficacy'' and introducing ill-understood standards of behavior shared generative AI spaces.
% fear of embarrassment and unclear standards of behavior for sharing generative AI systems.
\subsection{Tensions of Social Support}
Participants' expressed detailed concerns when it came to using generative AI for their small business, and how social use, even alongside trusted confidants, may not always be desirable, specifically pointing to reputational concerns, embarrassment and unclear social norms. 
To start, in response to S9 (see Appendix Figure ~\ref{fig:storyboard_9}), which featured two entrepreneurs from the same community sharing an interface into ChatGPT, P2 highlighted how shared use with another entrepreneur could intensify feelings of being left behind, especially if there was a large disparity in how comfortable the entrepreneurs felt using generative AI. 
% Despite the benefits of getting support while prompting, building self-efficacy, and discovering new ways to use the generative AI, that collaborative uses of generative AI technologies could foster, entrepreneurs and support personnel also reflected on the emergent tensions of these systems including 
% For instance, if one entrepreneur had significantly more experience with using generative AI than the other, then collaborating could intensify feelings of being left behind, and be \textit{``demoralizing''} for the entrepreneur with less experience.
This sentiment was echoed by E3 who said she would be scared to use a \textit{``stupid''} prompt that others might see, fretting her reputation as a competent business owner would be at stake. 
While E3 was not open to sharing prompts with others, she indicated that she would be interested to observe others' prompts. 
% At the same time, she questioned whether other entrepreneurs would see that as acceptable behavior given the lack of reciprocity.
To E2, simply observing would not be acceptable, saying \textit{``You don't want it being one person is doing all the work and the other person [is not]. [Then], you don't get anything out of it!''}. 

While entrepreneurs wrestled with these social norms of collaboratively using generative AI technologies (such as \textit{whether} to share prompts with others), E1, who shared a ChatGPT Premium account with business partners, had to figure out \textit{when} to share prompts with others.
He discussed how he actively chose to use his personal, free account, despite the lower performing model, when working on side projects to not \textit{``get anybody mad''} by adding \textit{``clutter''} to the account that he and business partners shared.
% Choosing \textit{whether} and \textit{when} to share prompts was a dimension entrepreneurs had to consider when reflecting upon storyboards or personal experience, and when entrepreneurs had different preferences on how they would like to use generative AI systems among others, disagreement could arise. 
Similarly \textit{who} to collaborate on a generative AI system with added another layer of tension. 
For instance, E2 reflected on how she would not want to share with a \textit{``techie person''} because it was important to her that she had similar \textit{``core values''} with the person that she would be sharing with. 
% Overall, when tensions could arise from collaborative uses of generative AI due unclear and dissonant expectations for behavior, heightened potential for embarrassment and degrading feelings of self-efficacy, entrepreneurs acknowledged that shared trust with others' was paramount to overcoming these challenges.
Entrepreneurs like E4 and E7 illustrated how social interactions within their networks influenced the formation of their perspectives and usage patterns of generative AI in their business practices.
For example, E4 acknowledged that her decisions about using ChatGPT was heavily influenced by her anticipation of her peers' opinions.
When reflecting on creating an artist statement, she shared how she would feel embarrassed if a peer asked her if she used ChatGPT to generate it, because she would want them to think she was \textit{``smart enough already to be able to write it from [her] own intellect and brain rather than [needing] a tool to help [her]''}.
% Ultimately, while E4 thought most people would encourage her to use the tool to help her convey her ideas, she decided she would not use ChatGPT to help her with her artist statement out of the fear of potential embarrassment associated with use.
On the other hand, E7 (also an artist entrepreneur) shared that, while she was eager to talk about image-generation technologies with other artists, her uncertainty around others' opinions made it difficult to have those conversations. 
She believed that, for art, as long as people were transparent about their use of generative AI technologies (specifically image-generation platforms), they should be allowed to use them as they wish. 
% Yet, she thinks she her viewpoint about how generative AI should be used in creative processes is an \textit{``unpopular opinion''} in artists circles. 
E7 indicated that talking about these technologies with peers is \textit{``a conversation that depending on [who] you're talking to [can] be a bit of a touchy subject''} and that unless E7 found other entrepreneurs excited to talk about generative AI, she held back.
In this way, her uncertainty around other entrepreneurs' stance in using generative AI, fueled by myriad controversies related to workforce roles, copyright, and data privacy, among others, made it difficult for her to console her network's opinions on the ethical aspects of different use cases.

\section{Limitations}
One limitation of this work was that participants tended to be the most engaged entrepreneurs at the entrepreneurial hubs we recruited from, potentially skewing the results to reflect more opportunistic relationships with generative AI (i.e., those who had less interest in generative AI may have been less likely to respond to our recruitment efforts).
While entrepreneurs and support personnel were keen to discuss the pitfalls of using generative AI for small business, future work can prioritize more critical stances and take up longitudinal approaches to observe use (and non-use) over the long term. 
Further, the small sample size of 11 participants limits the robustness of our findings.

\section{Discussion and Future Work}
From our exploratory study, we provided a snapshot for how small, offline networks facilitated entrepreneurs' adoption, continued use, and refusal of use of generative AI systems, as well as the tensions of this social support.
As the chat-based interaction paradigms of predominant generative AI technologies inevitably evolve in attempts to better support workers~\cite{morris2023design}, we consider how current trajectories of interface design may overlook the needs of workers with less traditional forms of employment, such as the entrepreneurs in our study.
% We consider new forms of work (independent workers such as entrepreneurs, crowd workers, online gig workers) may be overlooked (CITE). 
For instance, while a growing body of scholarship aims to address social use of generative AI tools in an organizational context~\cite{10.1145/3544549.3573802, 10.1145/3500868.3559450, brachman2024knowledge}---where workers have shared goals, teams and incentive structures---the entrepreneurs in our study pursued distinct goals, and worried about their reputations and intellectual property when seeking support from their offline networks (even among vetted networks). 
And yet, the entrepreneurs in our study resourcefully leveraged social support to overcome isolated work conditions and the difficulty of keeping pace with technological advancements, such as by repurposing single-user accounts for collaborative use (not unlike how crowd workers overcome isolation and complete tasks collaboratively \cite{crowdGray2016}. 

Looking towards future work, we consider theoretical frameworks that may be helpful when informing how generative AI systems could account for the benefits and tensions of social support, due to the nuanced social relationships between entrepreneurs and their support networks.
For instance, one relevant theory may be ``coopetition''~\cite{kuhn2015little}---or how social relationships between entrepreneurs and their networks comprise a combination of competitive and cooperative dynamics.
Concretely, generative AI systems for collaborative use among entrepreneurs could include more dynamic controls for information sharing, where a user could select which parts of their inputs and outputs are included or excluded from what other users see (also known as ``broadcast levels'' \cite{broadcastlevels}). 
By providing flexible visibility options, designers can aid entrepreneurs in navigating disparate preferences when engaging social support to use generative AI technology outside of often infrequent in-person support.
This could be particularly important in the case of ethics-based conversations between entrepreneurs, where one entrepreneurs' use of generative AI technology could be used to harm their reputation by another entrepreneur looking to gain a competitive edge.

\section{Conclusion}
% This paper investigated how local entrepreneurs utilize their local network to facilitate maintenance, repair, and refusal of use of generative AI technologies.
This paper investigated the role of social support---among vetted, offline networks---for small business owners integrating generative AI into their workflows.
To do so, we probed the benefits and tensions of entrepreneurs' current and preferred sources of social support. 
Despite being designed for single-user use, entrepreneurs shared the benefits of collaboratively using generative AI: discovering new use cases, developing self-efficacy, easing techno-anxieties, and wrap-around support. 
% We shared how collaboration around the use of generative AI aided entrepreneurs in discovering new use cases for the technology in their small business, develop self-efficacy, assuage techno-anxieties, and receive wrap-around support. 
Yet, even among the trusted and vetted networks they built, these configurations also facilitated social comparison, feelings of being left behind, reputational concerns, and surfaced unclear social norms during shared use of generative AI. 
We therefore consider how generative AI systems can be designed with these nuanced relationships---a combination of cooperation and competition---to support both use and non-use.

\bibliographystyle{ACM-Reference-Format}
\bibliography{sample-base}


\begin{thebibliography}{43}

%%% the Web address.

\ifx \showCODEN    \undefined \def \showCODEN     #1{\unskip}     \fi
\ifx \showDOI      \undefined \def \showDOI       #1{#1}\fi
\ifx \showISBNx    \undefined \def \showISBNx     #1{\unskip}     \fi
\ifx \showISBNxiii \undefined \def \showISBNxiii  #1{\unskip}     \fi
\ifx \showISSN     \undefined \def \showISSN      #1{\unskip}     \fi
\ifx \showLCCN     \undefined \def \showLCCN      #1{\unskip}     \fi
\ifx \shownote     \undefined \def \shownote      #1{#1}          \fi
\ifx \showarticletitle \undefined \def \showarticletitle #1{#1}   \fi
\ifx \showURL      \undefined \def \showURL       {\relax}        \fi
% invisible to TeX
\providecommand\bibfield[2]{#2}
\providecommand\bibinfo[2]{#2}
\providecommand\natexlab[1]{#1}
\providecommand\showeprint[2][]{arXiv:#2}

\bibitem[Avle et~al\mbox{.}(2019)]%
        {avle2019additional}
\bibfield{author}{\bibinfo{person}{Seyram Avle}, \bibinfo{person}{Julie Hui}, \bibinfo{person}{Silvia Lindtner}, {and} \bibinfo{person}{Tawanna Dillahunt}.} \bibinfo{year}{2019}\natexlab{}.
\newblock \showarticletitle{Additional Labors of the Entrepreneurial Self}.
\newblock \bibinfo{journal}{\emph{Proc. ACM Hum.-Comput. Interact.}} \bibinfo{volume}{3}, \bibinfo{number}{CSCW}, Article \bibinfo{articleno}{218} (\bibinfo{date}{nov} \bibinfo{year}{2019}), \bibinfo{numpages}{24}~pages.
\newblock
\urldef\tempurl%
\url{https://doi.org/10.1145/3359320}
\showDOI{\tempurl}


\bibitem[Brachman et~al\mbox{.}(2024)]%
        {brachman2024knowledge}
\bibfield{author}{\bibinfo{person}{Michelle Brachman}, \bibinfo{person}{Amina El-Ashry}, \bibinfo{person}{Casey Dugan}, {and} \bibinfo{person}{Werner Geyer}.} \bibinfo{year}{2024}\natexlab{}.
\newblock \showarticletitle{How Knowledge Workers Use and Want to Use LLMs in an Enterprise Context}. In \bibinfo{booktitle}{\emph{Extended Abstracts of the CHI Conference on Human Factors in Computing Systems}}. \bibinfo{pages}{1--8}.
\newblock


\bibitem[Charmaz(2006)]%
        {charmaz}
\bibfield{author}{\bibinfo{person}{Kathy Charmaz}.} \bibinfo{year}{2006}\natexlab{}.
\newblock \bibinfo{booktitle}{\emph{Constructing Grounded Theory: A Practical Guide Through Qualitative Analysis}}. Vol.~\bibinfo{volume}{1}.
\newblock \bibinfo{pages}{72--95}.
\newblock


\bibitem[Company(2023)]%
        {McKinseyPromptEngineering}
\bibfield{author}{\bibinfo{person}{McKinsey~\& Company}.} \bibinfo{year}{2023}\natexlab{}.
\newblock \showarticletitle{What is prompt engineering?}
\newblock \bibinfo{journal}{\emph{McKinsey Explainers}} (\bibinfo{year}{2023}).
\newblock
\urldef\tempurl%
\url{https://www.mckinsey.com/featured-insights/mckinsey-explainers/what-is-prompt-engineering}
\showURL{%
\tempurl}


\bibitem[Corden and Sainsbury(2006)]%
        {quoting-convention}
\bibfield{author}{\bibinfo{person}{Anne Corden} {and} \bibinfo{person}{Roy Sainsbury}.} \bibinfo{year}{2006}\natexlab{}.
\newblock \showarticletitle{Using Verbatim Quotations in Reporting Qualitative Social Research: The views of research users}.
\newblock  (\bibinfo{date}{01} \bibinfo{year}{2006}).
\newblock


\bibitem[Dillahunt et~al\mbox{.}(2018)]%
        {dillahunt2018entrepreneurship}
\bibfield{author}{\bibinfo{person}{Tawanna~R Dillahunt}, \bibinfo{person}{Vaishnav Kameswaran}, \bibinfo{person}{Desiree McLain}, \bibinfo{person}{Minnie Lester}, \bibinfo{person}{Delores Orr}, {and} \bibinfo{person}{Kentaro Toyama}.} \bibinfo{year}{2018}\natexlab{}.
\newblock \showarticletitle{Entrepreneurship and the socio-technical chasm in a lean economy}. In \bibinfo{booktitle}{\emph{Proceedings of the 2018 CHI Conference on Human Factors in Computing Systems}}. \bibinfo{pages}{1--14}.
\newblock


\bibitem[Dillahunt et~al\mbox{.}(2022)]%
        {10.1145/3491102.3501949}
\bibfield{author}{\bibinfo{person}{Tawanna~R Dillahunt}, \bibinfo{person}{Alex~Jiahong Lu}, \bibinfo{person}{Aarti Israni}, \bibinfo{person}{Ruchita Lodha}, \bibinfo{person}{Savana Brewer}, \bibinfo{person}{Tiera~S Robinson}, \bibinfo{person}{Angela~Brown Wilson}, {and} \bibinfo{person}{Earnest Wheeler}.} \bibinfo{year}{2022}\natexlab{}.
\newblock \showarticletitle{The Village: Infrastructuring Community-based Mentoring to Support Adults Experiencing Poverty}. In \bibinfo{booktitle}{\emph{Proceedings of the 2022 CHI Conference on Human Factors in Computing Systems}} (New Orleans, LA, USA) \emph{(\bibinfo{series}{CHI '22})}. \bibinfo{publisher}{Association for Computing Machinery}, \bibinfo{address}{New York, NY, USA}, Article \bibinfo{articleno}{574}, \bibinfo{numpages}{17}~pages.
\newblock
\showISBNx{9781450391573}
\urldef\tempurl%
\url{https://doi.org/10.1145/3491102.3501949}
\showDOI{\tempurl}


\bibitem[Garcia et~al\mbox{.}(2020)]%
        {garcia2020no}
\bibfield{author}{\bibinfo{person}{Patricia Garcia}, \bibinfo{person}{Tonia Sutherland}, \bibinfo{person}{Marika Cifor}, \bibinfo{person}{Anita~Say Chan}, \bibinfo{person}{Lauren Klein}, \bibinfo{person}{Catherine D'Ignazio}, {and} \bibinfo{person}{Niloufar Salehi}.} \bibinfo{year}{2020}\natexlab{}.
\newblock \showarticletitle{No: Critical refusal as feminist data practice}. In \bibinfo{booktitle}{\emph{conference companion publication of the 2020 on computer supported cooperative work and social computing}}. \bibinfo{pages}{199--202}.
\newblock


\bibitem[Gero et~al\mbox{.}(2022)]%
        {gero2022sparks}
\bibfield{author}{\bibinfo{person}{Katy~Ilonka Gero}, \bibinfo{person}{Vivian Liu}, {and} \bibinfo{person}{Lydia Chilton}.} \bibinfo{year}{2022}\natexlab{}.
\newblock \showarticletitle{Sparks: Inspiration for science writing using language models}. In \bibinfo{booktitle}{\emph{Proceedings of the 2022 ACM Designing Interactive Systems Conference}}. \bibinfo{pages}{1002--1019}.
\newblock


\bibitem[Golembewski and Selby(2010)]%
        {golembewski2010ideation}
\bibfield{author}{\bibinfo{person}{Michael Golembewski} {and} \bibinfo{person}{Mark Selby}.} \bibinfo{year}{2010}\natexlab{}.
\newblock \showarticletitle{Ideation decks: a card-based design ideation tool}. In \bibinfo{booktitle}{\emph{Proceedings of the 8th ACM Conference on Designing Interactive Systems}}. \bibinfo{pages}{89--92}.
\newblock


\bibitem[Gray et~al\mbox{.}(2016)]%
        {crowdGray2016}
\bibfield{author}{\bibinfo{person}{Mary~L. Gray}, \bibinfo{person}{Siddharth Suri}, \bibinfo{person}{Syed~Shoaib Ali}, {and} \bibinfo{person}{Deepti Kulkarni}.} \bibinfo{year}{2016}\natexlab{}.
\newblock \showarticletitle{The Crowd is a Collaborative Network}. In \bibinfo{booktitle}{\emph{Proceedings of the 19th ACM Conference on Computer-Supported Cooperative Work \& Social Computing}} (San Francisco, California, USA) \emph{(\bibinfo{series}{CSCW '16})}. \bibinfo{publisher}{Association for Computing Machinery}, \bibinfo{address}{New York, NY, USA}, \bibinfo{pages}{134–147}.
\newblock
\showISBNx{9781450335928}
\urldef\tempurl%
\url{https://doi.org/10.1145/2818048.2819942}
\showDOI{\tempurl}


\bibitem[Grevet and Gilbert(2015)]%
        {piggybackPrototyping}
\bibfield{author}{\bibinfo{person}{Catherine Grevet} {and} \bibinfo{person}{Eric Gilbert}.} \bibinfo{year}{2015}\natexlab{}.
\newblock \showarticletitle{Piggyback Prototyping: Using Existing, Large-Scale Social Computing Systems to Prototype New Ones}. In \bibinfo{booktitle}{\emph{Proceedings of the 33rd Annual ACM Conference on Human Factors in Computing Systems}} (Seoul, Republic of Korea) \emph{(\bibinfo{series}{CHI '15})}. \bibinfo{publisher}{Association for Computing Machinery}, \bibinfo{address}{New York, NY, USA}, \bibinfo{pages}{4047–4056}.
\newblock
\showISBNx{9781450331456}
\urldef\tempurl%
\url{https://doi.org/10.1145/2702123.2702395}
\showDOI{\tempurl}


\bibitem[Harrington et~al\mbox{.}(2019)]%
        {harrington2019deconstructing}
\bibfield{author}{\bibinfo{person}{Christina Harrington}, \bibinfo{person}{Sheena Erete}, {and} \bibinfo{person}{Anne~Marie Piper}.} \bibinfo{year}{2019}\natexlab{}.
\newblock \showarticletitle{Deconstructing community-based collaborative design: Towards more equitable participatory design engagements}.
\newblock \bibinfo{journal}{\emph{Proceedings of the ACM on Human-Computer Interaction}} \bibinfo{volume}{3}, \bibinfo{number}{CSCW} (\bibinfo{year}{2019}), \bibinfo{pages}{1--25}.
\newblock


\bibitem[Hauptman et~al\mbox{.}(2022)]%
        {10.1145/3500868.3559450}
\bibfield{author}{\bibinfo{person}{Allyson~I. Hauptman}, \bibinfo{person}{Wen Duan}, {and} \bibinfo{person}{Nathan~J. Mcneese}.} \bibinfo{year}{2022}\natexlab{}.
\newblock \showarticletitle{The Components of Trust for Collaborating With AI Colleagues}. In \bibinfo{booktitle}{\emph{Companion Publication of the 2022 Conference on Computer Supported Cooperative Work and Social Computing}} (Virtual Event, Taiwan) \emph{(\bibinfo{series}{CSCW'22 Companion})}. \bibinfo{publisher}{Association for Computing Machinery}, \bibinfo{address}{New York, NY, USA}, \bibinfo{pages}{72–75}.
\newblock
\showISBNx{9781450391900}
\urldef\tempurl%
\url{https://doi.org/10.1145/3500868.3559450}
\showDOI{\tempurl}


\bibitem[Hui et~al\mbox{.}(2020)]%
        {10.1145/3313831.3376363}
\bibfield{author}{\bibinfo{person}{Julie Hui}, \bibinfo{person}{Nefer~Ra Barber}, \bibinfo{person}{Wendy Casey}, \bibinfo{person}{Suzanne Cleage}, \bibinfo{person}{Danny~C. Dolley}, \bibinfo{person}{Frances Worthy}, \bibinfo{person}{Kentaro Toyama}, {and} \bibinfo{person}{Tawanna~R. Dillahunt}.} \bibinfo{year}{2020}\natexlab{}.
\newblock \showarticletitle{Community Collectives: Low-tech Social Support for Digitally-Engaged Entrepreneurship}. In \bibinfo{booktitle}{\emph{Proceedings of the 2020 CHI Conference on Human Factors in Computing Systems}} (, Honolulu, HI, USA,) \emph{(\bibinfo{series}{CHI '20})}. \bibinfo{publisher}{Association for Computing Machinery}, \bibinfo{address}{New York, NY, USA}, \bibinfo{pages}{1–15}.
\newblock
\showISBNx{9781450367080}
\urldef\tempurl%
\url{https://doi.org/10.1145/3313831.3376363}
\showDOI{\tempurl}


\bibitem[Hui et~al\mbox{.}(2018)]%
        {hui2018making}
\bibfield{author}{\bibinfo{person}{Julie Hui}, \bibinfo{person}{Kentaro Toyama}, \bibinfo{person}{Joyojeet Pal}, {and} \bibinfo{person}{Tawanna Dillahunt}.} \bibinfo{year}{2018}\natexlab{}.
\newblock \showarticletitle{Making a living my way: Necessity-driven entrepreneurship in resource-constrained communities}.
\newblock \bibinfo{journal}{\emph{Proceedings of the ACM on Human-Computer Interaction}} \bibinfo{volume}{2}, \bibinfo{number}{CSCW} (\bibinfo{year}{2018}), \bibinfo{pages}{1--24}.
\newblock


\bibitem[Kawakami et~al\mbox{.}(2023)]%
        {kawakami2023sensing}
\bibfield{author}{\bibinfo{person}{Anna Kawakami}, \bibinfo{person}{Shreya Chowdhary}, \bibinfo{person}{Shamsi~T Iqbal}, \bibinfo{person}{Q~Vera Liao}, \bibinfo{person}{Alexandra Olteanu}, \bibinfo{person}{Jina Suh}, {and} \bibinfo{person}{Koustuv Saha}.} \bibinfo{year}{2023}\natexlab{}.
\newblock \showarticletitle{Sensing wellbeing in the workplace, why and for whom? envisioning impacts with organizational stakeholders}.
\newblock \bibinfo{journal}{\emph{Proceedings of the ACM on Human-Computer Interaction}} \bibinfo{volume}{7}, \bibinfo{number}{CSCW2} (\bibinfo{year}{2023}), \bibinfo{pages}{1--33}.
\newblock


\bibitem[Kotturi et~al\mbox{.}(2024a)]%
        {kotturi2024deconstructing}
\bibfield{author}{\bibinfo{person}{Yasmine Kotturi}, \bibinfo{person}{Angel Anderson}, \bibinfo{person}{Glenn Ford}, \bibinfo{person}{Michael Skirpan}, {and} \bibinfo{person}{Jeffrey~P Bigham}.} \bibinfo{year}{2024}\natexlab{a}.
\newblock \showarticletitle{Deconstructing the Veneer of Simplicity: Co-Designing Introductory Generative AI Workshops with Local Entrepreneurs}. In \bibinfo{booktitle}{\emph{Proceedings of the CHI Conference on Human Factors in Computing Systems}}. \bibinfo{pages}{1--16}.
\newblock


\bibitem[Kotturi et~al\mbox{.}(2022)]%
        {10.1145/3491102.3517708}
\bibfield{author}{\bibinfo{person}{Yasmine Kotturi}, \bibinfo{person}{Herman~T Johnson}, \bibinfo{person}{Michael Skirpan}, \bibinfo{person}{Sarah~E Fox}, \bibinfo{person}{Jeffrey~P Bigham}, {and} \bibinfo{person}{Amy Pavel}.} \bibinfo{year}{2022}\natexlab{}.
\newblock \showarticletitle{Tech Help Desk: Support for Local Entrepreneurs Addressing the Long Tail of Computing Challenges}. In \bibinfo{booktitle}{\emph{Proceedings of the 2022 CHI Conference on Human Factors in Computing Systems}} (New Orleans, LA, USA,) \emph{(\bibinfo{series}{CHI '22})}. \bibinfo{publisher}{Association for Computing Machinery}, \bibinfo{address}{New York, NY, USA}, Article \bibinfo{articleno}{15}, \bibinfo{numpages}{15}~pages.
\newblock
\showISBNx{9781450391573}
\urldef\tempurl%
\url{https://doi.org/10.1145/3491102.3517708}
\showDOI{\tempurl}


\bibitem[Kotturi et~al\mbox{.}(2024b)]%
        {kotturi2024peerdea}
\bibfield{author}{\bibinfo{person}{Yasmine Kotturi}, \bibinfo{person}{Jenny Yu}, \bibinfo{person}{Pranav Khadpe}, \bibinfo{person}{Erin Gatz}, \bibinfo{person}{Harvey Zheng}, \bibinfo{person}{Sarah~E Fox}, {and} \bibinfo{person}{Chinmay Kulkarni}.} \bibinfo{year}{2024}\natexlab{b}.
\newblock \showarticletitle{Peerdea: Co-Designing a Peer Support Platform with Creative Entrepreneurs}.
\newblock \bibinfo{journal}{\emph{Proceedings of the ACM on Human-Computer Interaction}} \bibinfo{volume}{8}, \bibinfo{number}{CSCW1} (\bibinfo{year}{2024}), \bibinfo{pages}{1--24}.
\newblock


\bibitem[Kraus et~al\mbox{.}(2019)]%
        {digital-entrepreneurship}
\bibfield{author}{\bibinfo{person}{Sascha Kraus}, \bibinfo{person}{Carolin Palmer}, \bibinfo{person}{Norbert Kailer}, \bibinfo{person}{Lukas~K. Friedrich}, {and} \bibinfo{person}{Jonathan Spitzer}.} \bibinfo{year}{2019}\natexlab{}.
\newblock \showarticletitle{Digital entrepreneurship: A research agenda on new business models for the twenty-first century}.
\newblock \bibinfo{journal}{\emph{International Journal of Entrepreneurial Behaviour \& Research}} \bibinfo{volume}{25}, \bibinfo{number}{2} (\bibinfo{year}{2019}), \bibinfo{pages}{353--375}.
\newblock
\showISBNx{13552554}
\urldef\tempurl%
\url{https://www.proquest.com/scholarly-journals/digital-entrepreneurship/docview/2182418213/se-2}
\showURL{%
\tempurl}
\newblock
\shownote{Copyright - © Emerald Publishing Limited 2018; Last updated - 2023-11-26; SubjectsTermNotLitGenreText - Balearic Islands; United States--US; Finland; Spain}.


\bibitem[Kuhn and Galloway(2015)]%
        {kuhn2015little}
\bibfield{author}{\bibinfo{person}{Kristine~M Kuhn} {and} \bibinfo{person}{Tera~L Galloway}.} \bibinfo{year}{2015}\natexlab{}.
\newblock \showarticletitle{With a little help from my competitors: Peer networking among artisan entrepreneurs}.
\newblock \bibinfo{journal}{\emph{Entrepreneurship Theory and Practice}} \bibinfo{volume}{39}, \bibinfo{number}{3} (\bibinfo{year}{2015}), \bibinfo{pages}{571--600}.
\newblock


\bibitem[Le~Dantec and Fox(2015)]%
        {le2015strangers}
\bibfield{author}{\bibinfo{person}{Christopher~A Le~Dantec} {and} \bibinfo{person}{Sarah Fox}.} \bibinfo{year}{2015}\natexlab{}.
\newblock \showarticletitle{Strangers at the gate: Gaining access, building rapport, and co-constructing community-based research}. In \bibinfo{booktitle}{\emph{Proceedings of the 18th ACM conference on computer supported cooperative work \& social computing}}. \bibinfo{pages}{1348--1358}.
\newblock


\bibitem[Liu and Chilton(2022)]%
        {10.1145/3491102.3501825}
\bibfield{author}{\bibinfo{person}{Vivian Liu} {and} \bibinfo{person}{Lydia~B Chilton}.} \bibinfo{year}{2022}\natexlab{}.
\newblock \showarticletitle{Design Guidelines for Prompt Engineering Text-to-Image Generative Models}. In \bibinfo{booktitle}{\emph{Proceedings of the 2022 CHI Conference on Human Factors in Computing Systems}} (, New Orleans, LA, USA,) \emph{(\bibinfo{series}{CHI '22})}. \bibinfo{publisher}{Association for Computing Machinery}, \bibinfo{address}{New York, NY, USA}, Article \bibinfo{articleno}{384}, \bibinfo{numpages}{23}~pages.
\newblock
\showISBNx{9781450391573}
\urldef\tempurl%
\url{https://doi.org/10.1145/3491102.3501825}
\showDOI{\tempurl}


\bibitem[Marcoulides(1989)]%
        {technoAnxiety}
\bibfield{author}{\bibinfo{person}{George~A. Marcoulides}.} \bibinfo{year}{1989}\natexlab{}.
\newblock \showarticletitle{Measuring Computer Anxiety: The Computer Anxiety Scale}.
\newblock \bibinfo{journal}{\emph{Educational and Psychological Measurement}} \bibinfo{volume}{49}, \bibinfo{number}{3} (\bibinfo{year}{1989}), \bibinfo{pages}{733--739}.
\newblock
\urldef\tempurl%
\url{https://doi.org/10.1177/001316448904900328}
\showDOI{\tempurl}
\showeprint{https://doi.org/10.1177/001316448904900328}


\bibitem[{Michael M. Grynbaum and Ryan Mac}(2023)]%
        {nytimes2023Sue}
\bibfield{author}{\bibinfo{person}{{Michael M. Grynbaum and Ryan Mac}}.} \bibinfo{year}{2023}\natexlab{}.
\newblock \showarticletitle{New York Times sues OpenAI and Microsoft over A.I. Use of Copyrighted Work}.
\newblock \bibinfo{journal}{\emph{The New York Times}} (\bibinfo{year}{2023}).
\newblock
\urldef\tempurl%
\url{https://www.nytimes.com/2023/12/27/business/media/new-york-times-open-ai-microsoft-lawsuit.html}
\showURL{%
\tempurl}


\bibitem[{MIT Sloan}(2023)]%
        {mitsloan-generative-ai}
\bibfield{author}{\bibinfo{person}{{MIT Sloan}}.} \bibinfo{year}{2023}\natexlab{}.
\newblock \bibinfo{title}{Legal Issues Presented by Generative AI}.
\newblock \bibinfo{howpublished}{\url{https://mitsloan.mit.edu/ideas-made-to-matter/legal-issues-presented-generative-ai}}.
\newblock
\newblock
\shownote{Accessed: 2024-15-01}.


\bibitem[Morris et~al\mbox{.}(2023)]%
        {morris2023design}
\bibfield{author}{\bibinfo{person}{Meredith~Ringel Morris}, \bibinfo{person}{Carrie~J. Cai}, \bibinfo{person}{Jess Holbrook}, \bibinfo{person}{Chinmay Kulkarni}, {and} \bibinfo{person}{Michael Terry}.} \bibinfo{year}{2023}\natexlab{}.
\newblock \bibinfo{title}{The Design Space of Generative Models}.
\newblock
\newblock
\showeprint[arxiv]{2304.10547}~[cs.AI]


\bibitem[Oeldorf-Hirsch and Gergle(2020)]%
        {broadcastlevels}
\bibfield{author}{\bibinfo{person}{Anne Oeldorf-Hirsch} {and} \bibinfo{person}{Darren Gergle}.} \bibinfo{year}{2020}\natexlab{}.
\newblock \showarticletitle{'Who Knows What': Audience Targeting for Question Asking on Facebook}.
\newblock \bibinfo{journal}{\emph{Proc. ACM Hum.-Comput. Interact.}} \bibinfo{volume}{4}, \bibinfo{number}{GROUP}, Article \bibinfo{articleno}{11} (\bibinfo{date}{jan} \bibinfo{year}{2020}), \bibinfo{numpages}{20}~pages.
\newblock
\urldef\tempurl%
\url{https://doi.org/10.1145/3375191}
\showDOI{\tempurl}


\bibitem[OpenAI(2024)]%
        {openai_chatgpt}
\bibfield{author}{\bibinfo{person}{OpenAI}.} \bibinfo{year}{2024}\natexlab{}.
\newblock \bibinfo{title}{ChatGPT}.
\newblock \bibinfo{howpublished}{\url{https://openai.com/chatgpt/}}.
\newblock
\newblock
\shownote{Accessed: 2024-05-22}.


\bibitem[Otis et~al\mbox{.}(2023)]%
        {Otis2023UnevenImpact}
\bibfield{author}{\bibinfo{person}{Nicholas~G. Otis}, \bibinfo{person}{Rowan~P. Clarke}, \bibinfo{person}{Solene Delecourt}, \bibinfo{person}{David Holtz}, {and} \bibinfo{person}{Rembrand Koning}.} \bibinfo{year}{2023}\natexlab{}.
\newblock \bibinfo{title}{The Uneven Impact of Generative AI on Entrepreneurial Performance}.
\newblock \bibinfo{howpublished}{OSF Preprints}.
\newblock
\urldef\tempurl%
\url{https://doi.org/10.31219/osf.io/hdjpk}
\showDOI{\tempurl}


\bibitem[Pei and Crooks(2020)]%
        {10.1145/3313831.3376587}
\bibfield{author}{\bibinfo{person}{Lucy Pei} {and} \bibinfo{person}{Roderic Crooks}.} \bibinfo{year}{2020}\natexlab{}.
\newblock \showarticletitle{Attenuated Access: Accounting for Startup, Maintenance, and Affective Costs in Resource-Constrained Communities}. In \bibinfo{booktitle}{\emph{Proceedings of the 2020 CHI Conference on Human Factors in Computing Systems}} (, Honolulu, HI, USA,) \emph{(\bibinfo{series}{CHI '20})}. \bibinfo{publisher}{Association for Computing Machinery}, \bibinfo{address}{New York, NY, USA}, \bibinfo{pages}{1–15}.
\newblock
\showISBNx{9781450367080}
\urldef\tempurl%
\url{https://doi.org/10.1145/3313831.3376587}
\showDOI{\tempurl}


\bibitem[Roberts(1991)]%
        {roberts1991entrepreneurs}
\bibfield{author}{\bibinfo{person}{Edward~B Roberts}.} \bibinfo{year}{1991}\natexlab{}.
\newblock \bibinfo{booktitle}{\emph{Entrepreneurs in high technology: Lessons from MIT and beyond}}.
\newblock \bibinfo{publisher}{Oxford University Press}.
\newblock


\bibitem[Shin et~al\mbox{.}(2023)]%
        {10.1145/3544549.3573802}
\bibfield{author}{\bibinfo{person}{Joon~Gi Shin}, \bibinfo{person}{Janin Koch}, \bibinfo{person}{Andr\'{e}s Lucero}, \bibinfo{person}{Peter Dalsgaard}, {and} \bibinfo{person}{Wendy~E. Mackay}.} \bibinfo{year}{2023}\natexlab{}.
\newblock \showarticletitle{Integrating AI in Human-Human Collaborative Ideation}. In \bibinfo{booktitle}{\emph{Extended Abstracts of the 2023 CHI Conference on Human Factors in Computing Systems}} (Hamburg, Germany) \emph{(\bibinfo{series}{CHI EA '23})}. \bibinfo{publisher}{Association for Computing Machinery}, \bibinfo{address}{New York, NY, USA}, Article \bibinfo{articleno}{355}, \bibinfo{numpages}{5}~pages.
\newblock
\showISBNx{9781450394222}
\urldef\tempurl%
\url{https://doi.org/10.1145/3544549.3573802}
\showDOI{\tempurl}


\bibitem[Staff(2023)]%
        {IBMGenerativeAI}
\bibfield{author}{\bibinfo{person}{IBM~Research Staff}.} \bibinfo{year}{2023}\natexlab{}.
\newblock \showarticletitle{What is Generative AI?}
\newblock \bibinfo{journal}{\emph{IBM Research Blog}} (\bibinfo{year}{2023}).
\newblock
\urldef\tempurl%
\url{https://research.ibm.com/blog/what-is-generative-AI}
\showURL{%
\tempurl}


\bibitem[Strobelt et~al\mbox{.}(2023)]%
        {9908590}
\bibfield{author}{\bibinfo{person}{Hendrik Strobelt}, \bibinfo{person}{Albert Webson}, \bibinfo{person}{Victor Sanh}, \bibinfo{person}{Benjamin Hoover}, \bibinfo{person}{Johanna Beyer}, \bibinfo{person}{Hanspeter Pfister}, {and} \bibinfo{person}{Alexander~M. Rush}.} \bibinfo{year}{2023}\natexlab{}.
\newblock \showarticletitle{Interactive and Visual Prompt Engineering for Ad-hoc Task Adaptation with Large Language Models}.
\newblock \bibinfo{journal}{\emph{IEEE Transactions on Visualization and Computer Graphics}} \bibinfo{volume}{29}, \bibinfo{number}{1} (\bibinfo{year}{2023}), \bibinfo{pages}{1146--1156}.
\newblock
\urldef\tempurl%
\url{https://doi.org/10.1109/TVCG.2022.3209479}
\showDOI{\tempurl}


\bibitem[to~Visual(2023)]%
        {zq00icc13I4}
\bibfield{author}{\bibinfo{person}{Verbal to Visual}.} \bibinfo{year}{2023}\natexlab{}.
\newblock \bibinfo{title}{How to Draw Star People}.
\newblock \bibinfo{howpublished}{\url{https://youtu.be/zq00icc13I4?si=09FoEogESJg6O9OE}}.
\newblock
\newblock
\shownote{Accessed: 2024-01-23}.


\bibitem[Tohidi et~al\mbox{.}(2006)]%
        {tohidi2006getting}
\bibfield{author}{\bibinfo{person}{Maryam Tohidi}, \bibinfo{person}{William Buxton}, \bibinfo{person}{Ronald Baecker}, {and} \bibinfo{person}{Abigail Sellen}.} \bibinfo{year}{2006}\natexlab{}.
\newblock \showarticletitle{Getting the right design and the design right}. In \bibinfo{booktitle}{\emph{Proceedings of the SIGCHI conference on Human Factors in computing systems}}. \bibinfo{pages}{1243--1252}.
\newblock


\bibitem[Truong et~al\mbox{.}(2006)]%
        {storyboardingPractices}
\bibfield{author}{\bibinfo{person}{Khai Truong}, \bibinfo{person}{Gillian Hayes}, {and} \bibinfo{person}{Gregory Abowd}.} \bibinfo{year}{2006}\natexlab{}.
\newblock \showarticletitle{Storyboarding: an empirical determination of best practices and effective guidelines}. \bibinfo{pages}{12--21}.
\newblock
\urldef\tempurl%
\url{https://doi.org/10.1145/1142405.1142410}
\showDOI{\tempurl}


\bibitem[Verheijden and Funk(2023)]%
        {10.1145/3544549.3585680}
\bibfield{author}{\bibinfo{person}{Mathias~Peter Verheijden} {and} \bibinfo{person}{Mathias Funk}.} \bibinfo{year}{2023}\natexlab{}.
\newblock \showarticletitle{Collaborative Diffusion: Boosting Designerly Co-Creation with Generative AI}. In \bibinfo{booktitle}{\emph{Extended Abstracts of the 2023 CHI Conference on Human Factors in Computing Systems}} (, Hamburg, Germany,) \emph{(\bibinfo{series}{CHI EA '23})}. \bibinfo{publisher}{Association for Computing Machinery}, \bibinfo{address}{New York, NY, USA}, Article \bibinfo{articleno}{73}, \bibinfo{numpages}{8}~pages.
\newblock
\showISBNx{9781450394222}
\urldef\tempurl%
\url{https://doi.org/10.1145/3544549.3585680}
\showDOI{\tempurl}


\bibitem[von Hippel(2005)]%
        {10.7551/mitpress/2333.001.0001}
\bibfield{author}{\bibinfo{person}{Eric von Hippel}.} \bibinfo{year}{2005}\natexlab{}.
\newblock \bibinfo{booktitle}{\emph{{Democratizing Innovation}}}.
\newblock \bibinfo{publisher}{The MIT Press}.
\newblock
\showISBNx{9780262285636}
\urldef\tempurl%
\url{https://doi.org/10.7551/mitpress/2333.001.0001}
\showDOI{\tempurl}
\showeprint{https://direct.mit.edu/book-pdf/2092624/book\_9780262285636.pdf}


\bibitem[Wu et~al\mbox{.}(2023)]%
        {10.1145/3581641.3584059}
\bibfield{author}{\bibinfo{person}{Sherry Wu}, \bibinfo{person}{Hua Shen}, \bibinfo{person}{Daniel~S Weld}, \bibinfo{person}{Jeffrey Heer}, {and} \bibinfo{person}{Marco~Tulio Ribeiro}.} \bibinfo{year}{2023}\natexlab{}.
\newblock \showarticletitle{ScatterShot: Interactive In-context Example Curation for Text Transformation}. In \bibinfo{booktitle}{\emph{Proceedings of the 28th International Conference on Intelligent User Interfaces}} (Sydney, NSW, Australia) \emph{(\bibinfo{series}{IUI '23})}. \bibinfo{publisher}{Association for Computing Machinery}, \bibinfo{address}{New York, NY, USA}, \bibinfo{pages}{353–367}.
\newblock
\showISBNx{9798400701061}
\urldef\tempurl%
\url{https://doi.org/10.1145/3581641.3584059}
\showDOI{\tempurl}


\bibitem[Zamfirescu-Pereira et~al\mbox{.}(2023)]%
        {10.1145/3544548.3581388}
\bibfield{author}{\bibinfo{person}{J.D. Zamfirescu-Pereira}, \bibinfo{person}{Richmond~Y. Wong}, \bibinfo{person}{Bjoern Hartmann}, {and} \bibinfo{person}{Qian Yang}.} \bibinfo{year}{2023}\natexlab{}.
\newblock \showarticletitle{Why Johnny Can’t Prompt: How Non-AI Experts Try (and Fail) to Design LLM Prompts}. In \bibinfo{booktitle}{\emph{Proceedings of the 2023 CHI Conference on Human Factors in Computing Systems}} (, Hamburg, Germany,) \emph{(\bibinfo{series}{CHI '23})}. \bibinfo{publisher}{Association for Computing Machinery}, \bibinfo{address}{New York, NY, USA}, Article \bibinfo{articleno}{437}, \bibinfo{numpages}{21}~pages.
\newblock
\showISBNx{9781450394215}
\urldef\tempurl%
\url{https://doi.org/10.1145/3544548.3581388}
\showDOI{\tempurl}


\end{thebibliography}

\appendix
\label{appendix:A}
\section{Example Storyboards}
We include a subset of storyboards from our study which probed how offline networks---such as from local community centers~\cite{10.1145/3491102.3517708, 10.1145/3313831.3376363}---among entrepreneurs may provide social support online when in-person meetings were not possible. To improve readability, we overlay text on the storyboards' panels. Characters featured in the storyboards were drawn as ``star people'' \cite{zq00icc13I4} in attempts to avoid reinforcing gender and cultural stereotypes, and technical jargon was kept to a minimum. The paper format of storyboards helped to convey malleability of the ideas and foster critical conversations with participants~\cite{tohidi2006getting}.

% \FloatBarrier

% Storyboard 9
\begin{figure}[H]
\centering
\begin{minipage}{0.45\textwidth}
  \centering
  \includegraphics[width=\textwidth]{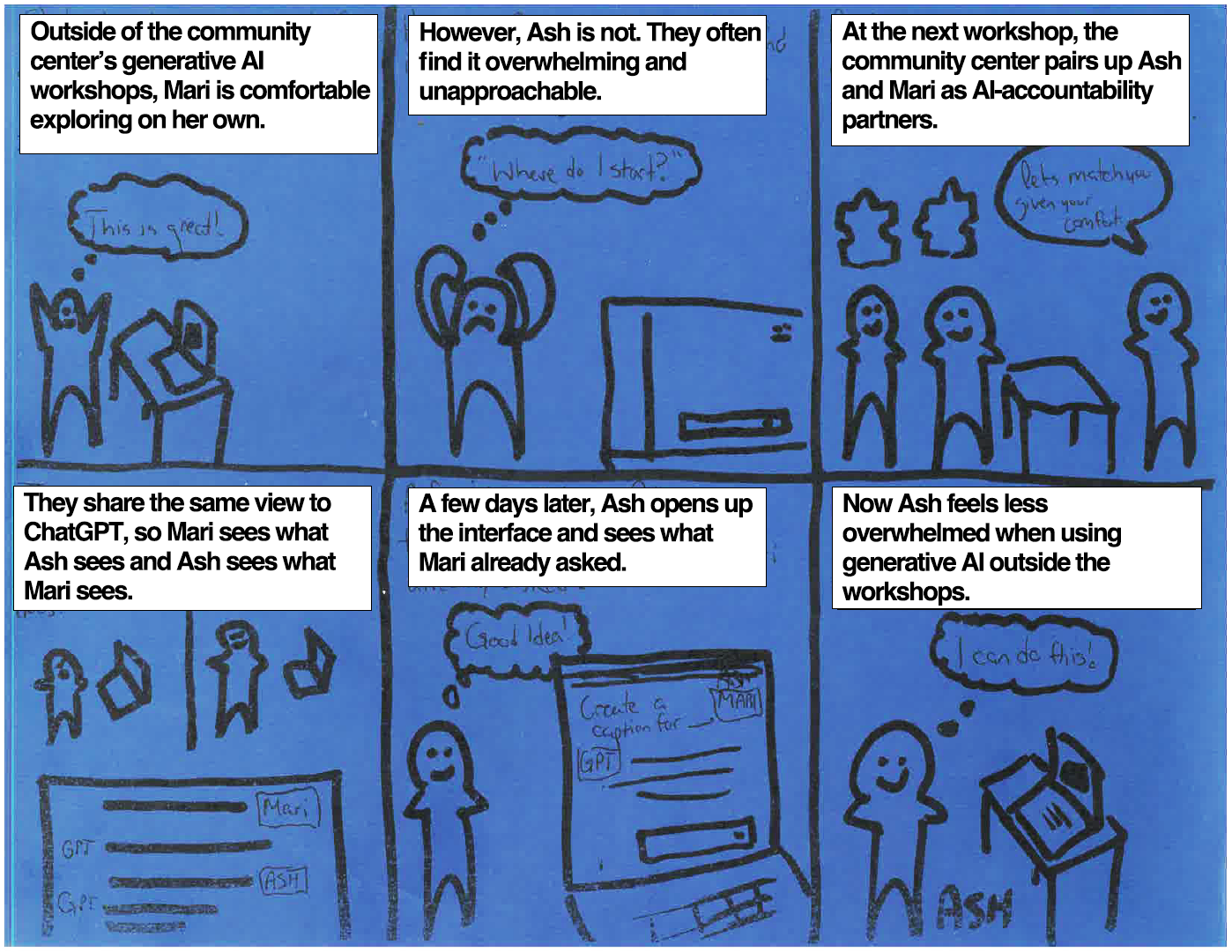}
  \label{fig:appendixImage1}
\end{minipage}
\hspace{0.05\textwidth}
\caption{Storyboard 9 probed how entrepreneurs may provide each other with informal social support alongside more formal resources~\cite{kotturi2024deconstructing}. In this storyboard, vetted peers who met at a local community center use a browser and chat-based generative AI platform collaboratively---when they cannot make it to the center---by seeing and building off of each others' prompts, and holding each other accountable to continued support.}
% \description{TODO}
\label{fig:storyboard_9}
\end{figure}

% Storyboard 17
\begin{figure}[H]
\centering
\begin{minipage}{0.45\textwidth}
  \centering
  \includegraphics[width=\textwidth]{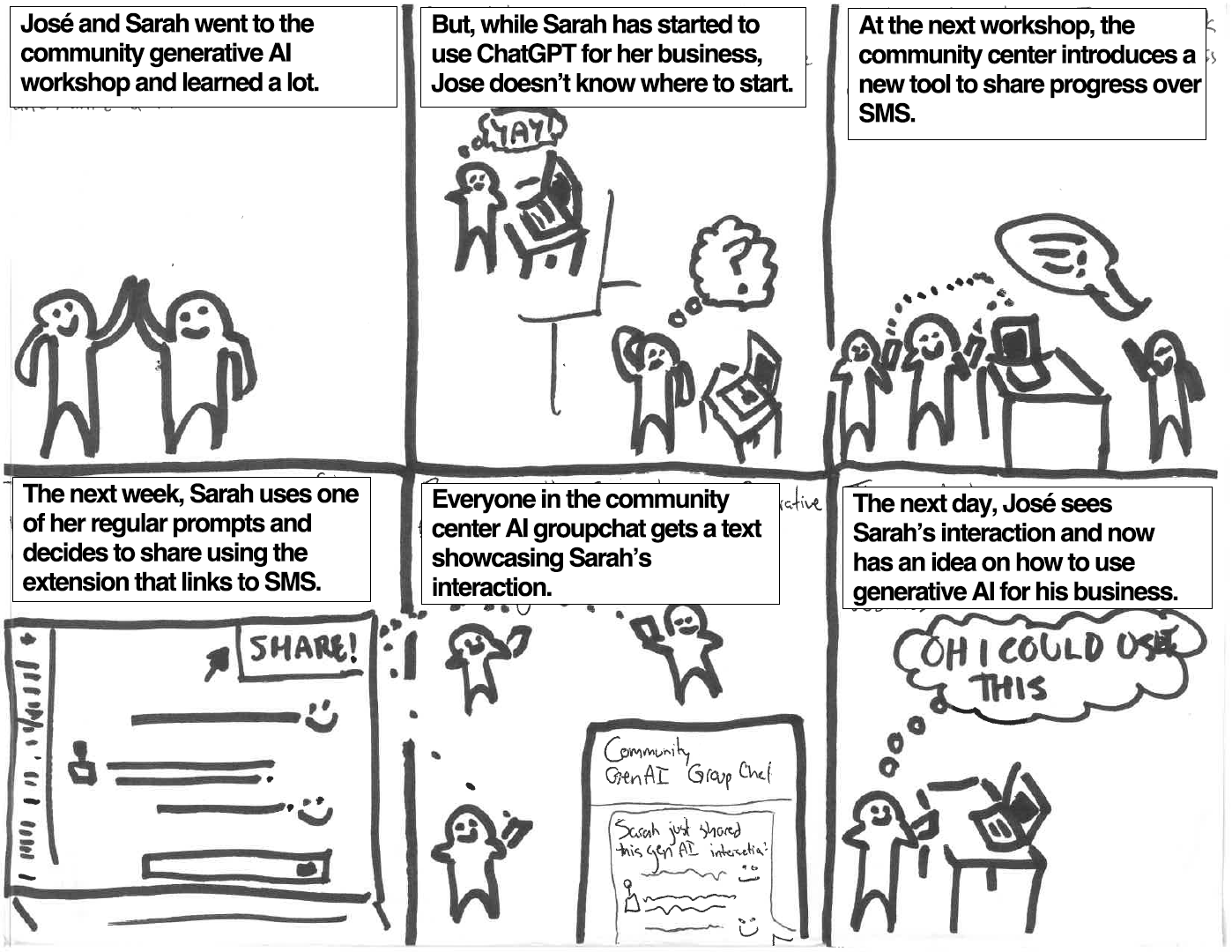}
  \label{fig:appendixImage1}
\end{minipage}
\hspace{0.05\textwidth}
% \vspace{-.78cm}
\caption{Storyboard 17 probed how a SMS-based system for social use of generative AI may meet entrepreneurs where are by leveraging an existing technology~\cite{piggybackPrototyping, 10.1145/3491102.3517708} to more readily share helpful prompts with other entrepreneurs. Similarly to Storyboard 9, this storyboard considers how entrepreneurs can support each other when they are unable to make it in person to their community center.}
\label{fig:storyboard_17}

\end{figure}

% Storyboard 24
% \begin{figure}[H]
% \centering
% \begin{minipage}{0.45\textwidth}
%   \centering
%   \includegraphics[width=\textwidth]{Appendix Content/storyboard_24.jpeg}
%   \label{fig:appendixImage1}
% \end{minipage}
% \hspace{0.05\textwidth}
% \vspace{-.78cm}
% \caption{Storyboard 24. As detailed in prior work, entrepreneurs were wary of seeking social support from unknown peers online}
% \label{fig:storyboards}
% \end{figure}

% % Storyboard 25
% \begin{figure}[H]
% \centering
% \begin{minipage}{0.45\textwidth}
%   \centering
%   \includegraphics[width=\textwidth]{Appendix Content/storyboard_25.png}
%   \label{fig:appendixImage1}
% \end{minipage}
% \hspace{0.05\textwidth}
% % \vspace{-.78cm}
% \caption{Storyboard 25 considered how entrepreneurs can engage their local networks, specifically close peers and friends, to collectively navigate concerns they may when using generative AI for their small business. In this storyboard, users leverage a browser-based platform to discuss ethical concerns when using image-generation technologies, and in doing so, learn from each others' perspectives.}
% \label{fig:storyboard_25}
% \end{figure}

% Storyboard 26
\begin{figure}[H]
\centering
\begin{minipage}{0.45\textwidth}
  \centering
  \includegraphics[width=\textwidth]{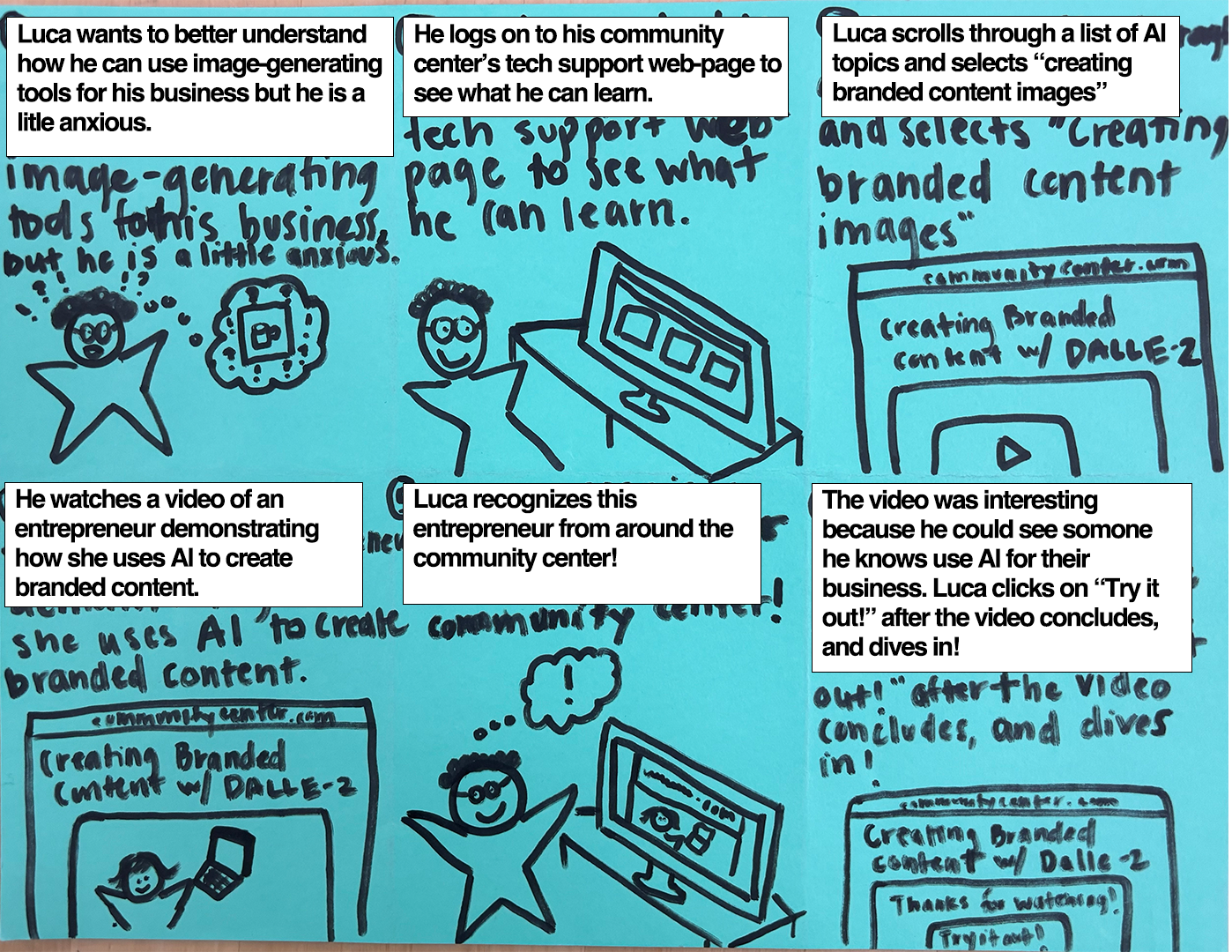}
  \label{fig:appendixImage1}
\end{minipage}
\hspace{0.05\textwidth}
\caption{Storyboard 26 probed how entrepreneurs can learn about integrating generative AI into their business workflows by learning from peers in their local community center, specifically through watching demonstrative videos by those in their community.}
\label{fig:storyboard_26}
\end{figure}

\end{document}